\documentclass[sigconf]{acmart}

\usepackage{bm}
\usepackage{amsmath}
\usepackage{amsfonts}
\usepackage{amssymb}
\usepackage{amsthm}
\usepackage[linesnumbered,vlined,ruled]{algorithm2e}
\usepackage{multirow}
\usepackage{caption}
\usepackage{subcaption}
\usepackage{etoolbox}
\makeatletter
\patchcmd{\@algocf@start}{%
  \begin{lrbox}{\algocf@algobox}%
}{%
  \rule{0.\textwidth}{\z@}%
  \begin{lrbox}{\algocf@algobox}%
  \begin{minipage}{0.5\textwidth}%
}{}{}
\patchcmd{\@algocf@finish}{%
  \end{lrbox}%
}{%
  \end{minipage}%
  \end{lrbox}%
}{}{}
\makeatother

\copyrightyear{2019}
\acmYear{2019} 
\setcopyright{none}
\acmConference[WWW '19]{Proceedings of the 2019 World Wide Web Conference}{May 13--17, 2019}{San Francisco, CA, USA}
\acmBooktitle{Proceedings of the 2019 World Wide Web Conference (WWW '19), May 13--17, 2019, San Francisco, CA, USA}
\acmPrice{}
\acmDOI{10.1145/3308558.3313417}
\acmISBN{978-1-4503-6674-8/19/05}

\begin{document}
 
\title{Knowledge Graph Convolutional Networks for Recommender Systems}

\author{Hongwei Wang}
\affiliation{Shanghai Jiao Tong University\\Shanghai, China}
\email{wanghongwei55@gmail.com}
\authornote{This work is done during the author's internship at Microsoft Research Asia.}

\author{Miao Zhao}
\affiliation{The Hong Kong Polytechnic University, Hong Kong, China}
\email{csmiaozhao@comp.polyu.edu.hk}

\author{Xing Xie}
\affiliation{Microsoft Research Asia\\Beijing, China}
\email{xingx@microsoft.com}

\author{Wenjie Li}
\affiliation{The Hong Kong Polytechnic University, Hong Kong, China}
\email{cswjli@comp.polyu.edu.hk}

\author{Minyi Guo}
\affiliation{Shanghai Jiao Tong University\\Shanghai, China}
\email{guo-my@cs.sjtu.edu.cn}
\authornote{Minyi Guo is the corresponding author. This work was partially sponsored by the National Basic Research 973 Program of China (2015CB352403) and National Natural Science Foundation of China (61272291).}

\begin{abstract}
	To alleviate sparsity and cold start problem of collaborative filtering based recommender systems, researchers and engineers usually collect attributes of users and items, and design delicate algorithms to exploit these additional information.
	In general, the attributes are not isolated but connected with each other, which forms a knowledge graph (KG).
	In this paper, we propose \textbf{K}nowledge \textbf{G}raph \textbf{C}onvolutional \textbf{N}etworks (KGCN), an end-to-end framework that captures inter-item relatedness effectively by mining their associated attributes on the KG.
	To automatically discover both high-order structure information and semantic information of the KG, we sample from the neighbors for each entity in the KG as their receptive field, then combine neighborhood information with bias when calculating the representation of a given entity.
	The receptive field can be extended to multiple hops away to model high-order proximity information and capture users' potential long-distance interests.
	Moreover, we implement the proposed KGCN in a minibatch fashion, which enables our model to operate on large datasets and KGs.
	We apply the proposed model to three datasets about movie, book, and music recommendation, and experiment results demonstrate that our approach outperforms strong recommender baselines.
\end{abstract}


\maketitle

\section{Introduction}
	With the advance of Internet technology, people can access a vast amount of online content, such as news \cite{zheng2018drn}, movies \cite{diao2014jointly}, and commodities \cite{zhou2018deep}.
	A notorious problem with online platforms is that the volume of items can be overwhelming to users.
	To alleviate the impact of information overloading, recommender systems (RS) is proposed to search for and recommend a small set of items to meet users' personalized interests. 

	A traditional recommendation technique is collaborative filtering (CF), which assigns users and items ID-based representation vectors, then models their interactions by specific operation such as inner product \cite{wang2017joint} or neural networks \cite{he2017neural}.
	However, CF-based methods usually suffer from sparsity of user-item interactions and the cold start problem.
	To address these limitations, researchers usually turn to feature-rich scenarios, where attributes of users and items are used to compensate for the sparsity and improve the performance of recommendation \cite{cheng2016wide,wang2018shine}.
	
	A few recent studies \cite{yu2014personalized,zhang2016collaborative,zhao2017meta,wang2018dkn,huang2018improving,wang2018ripple} have gone a step further than simply using attributes:
	They point out that attributes are not isolated but linked up with each other, which forms a \textit{knowledge graph} (KG).
	Typically, a KG is a directed heterogeneous graph in which nodes correspond to \textit{entities} (items or item attributes) and edges correspond to \textit{relations}.
	Compared with KG-free methods, incorporating KG into recommendation benefits the results in three ways \cite{wang2018ripple}:
	(1) The rich semantic relatedness among items in a KG can help explore their latent connections and improve the \textit{precision} of results;
	(2) The various types of relations in a KG are helpful for extending a user's interests reasonably and increasing the \textit{diversity} of recommended items;
	(3) KG connects a user's historically-liked and recommended items, thereby bringing \textit{explainability} to recommender systems.
	
	Despite the above benefits, utilizing KG in RS is rather challenging due to its high dimensionality and heterogeneity.
	One feasible way is to preprocess the KG by \textit{knowledge graph embedding} (KGE) methods \cite{wang2017knowledge}, which map entities and relations to low-dimensional representation vectors \cite{zhang2016collaborative,wang2018dkn,huang2018improving}.
	However, commonly-used KGE methods focus on modeling rigorous semantic relatedness (e.g., TransE \cite{bordes2013translating} and TransR \cite{lin2015learning} assume $head + relation = tail$), which are more suitable for in-graph applications such as KG completion and link prediction rather than recommendation.
	A more natural and intuitive way is to design a graph algorithm directly to exploit the KG structure \cite{yu2014personalized,zhao2017meta,wang2018ripple}.
	For example, PER \cite{yu2014personalized} and FMG \cite{zhao2017meta} treat KG as a heterogeneous information network, and extract meta-path/meta-graph based latent features to represent the connectivity between users and items along different types of relation paths/graphs.
	However, PER and FMG rely heavily on manually designed meta-paths or meta-graphs, which are hardly to be optimal in reality.
	RippleNet \cite{wang2018ripple} is a memory-network-like model that propagates users' potential preferences in the KG and explores their hierarchical interests.
	But note that the importance of relations is weakly characterized in RippleNet, because the embedding matrix of a relation $\bf R$ can hardly be trained to capture the sense of importance in the quadratic form ${\bf v}^\top {\bf R} {\bf h}$ ($\bf v$ and $\bf h$ are embedding vectors of two entities).
	In addition, the size of ripple set may go unpredictably with the increase of the size of KG, which incurs heavy computation and storage overhead.
	
	In this paper, we investigate the problem of KG-aware recommendation.
	Our design objective is to automatically capture both high-order structure and semantic information in the KG.	Inspired by graph convolutional networks (GCN)\footnote{We will revisit GCN in related work.} that try to generalize convolution to the graph domain, we propose \textbf{K}nowledge \textbf{G}raph \textbf{C}onvolutional \textbf{N}etworks (KGCN) for recommender systems.
	The key idea of KGCN is to aggregate and incorporate neighborhood information with bias when calculating the representation of a given entity in the KG.
	Such a design has two advantages:
	(1) Through the operation of neighborhood aggregation, the \textit{local proximity structure} is successfully captured and stored in each entity.
	(2) Neighbors are weighted by scores dependent on the connecting relation and specific user, which characterizes both the \textit{semantic information of KG} and \textit{users' personalized interests in relations}.
	Note that the size of an entity's neighbors varies and may be prohibitively large in the worst case.
	Therefore, we sample a fixed-size neighborhood of each node as the receptive field, which makes the cost of KGCN predictable.
	The definition of neighborhood for a given entity can also be extended hierarchically to multiple hops away to model \textit{high-order} entity dependencies and capture users' potential \textit{long-distance} interests.
	
	Empirically, we apply KGCN to three datasets: MovieLens-20M (movie), Book-Crossing (book), and Last.FM (music).
	The experiment results show that KGCN achieves average AUC gains of $4.4\%$, $8.1\%$, and $6.2\%$ in movie, book, and music recommendations, respectively, compared with state-of-the-art baselines for recommendation.
	
	Our contribution in this paper are summarized as follows:
	\begin{itemize}
		\item We propose knowledge graph convolutional networks, an end-to-end framework that explores users' preferences on the knowledge graph for recommender systems. By extending the receptive field of each entity in the KG, KGCN is able to capture users' high-order personalized interests.
		\item We conduct experiments on three real-world recommendation scenarios. The results demonstrate the efficacy of KGCN-LS over state-of-the-art baselines.
		\item We release the code of KGCN and datasets (knowledge graphs) to researchers for validating the reported results and conducting further research. The code and the data are available at \url{https://github.com/hwwang55/KGCN}.
	\end{itemize}

\section{Related Work}
	Our method is conceptually inspired by GCN.
	In general, GCN can be categorized as spectral methods and non-spectral methods.
	Spectral methods represent graphs and perform convolution in the spectral space.
	For example, Bruna et al. \cite{bruna2014spectral} define the convolution in Fourier domain and calculates the eigendecomposition of the graph Laplacian,
	Defferrard et al. \cite{defferrard2016convolutional} approximate the convolutional filters by Chebyshev expansion of the graph Laplacian,
	and Kipf et al. \cite{kipf2017semi} propose a convolutional architecture via a localized first-order approximation of spectral graph convolutions.
	In contrast, non-spectral methods operate on the original graph directly and define convolution for groups of nodes.
	To handle the neighborhoods with varying size and maintain the weight sharing property of CNN, researchers propose learning a weight matrix for each node degree \cite{duvenaud2015convolutional}, extracting locally connected regions from graphs \cite{niepert2016learning}, or sampling a fixed-size set of neighbors as the support size \cite{hamilton2017inductive}.
	Our work can be seen as a non-spectral method for a special type of graphs (i.e., knowledge graph).
	
	Our method also connects to PinSage \cite{ying2018graph} and GAT \cite{velickovic2017graph}.
	But note that both PinSage and GAT are designed for homogeneous graphs.
	The major difference between our work and the literature is that we offer a new perspective for recommender systems with the assistance of a heterogeneous knowledge graph.

\section{Knowledge Graph Convolutional Networks}
	In this section, we introduce the proposed KGCN model.
	We first formulate the knowledge-graph-aware recommendation problem.
	Then we present the design of a single layer of KGCN.
	At last, we introduce the complete learning algorithm for KGCN, as well as its minibatach implementation.
	
	\subsection{Problem Formulation}
		We formulate the knowledge-graph-aware recommendation problem as follows.
		In a typical recommendation scenario, we have a set of $M$ users $\mathcal U = \{u_1, u_2, ..., u_M\}$ and a set of $N$ items $\mathcal V = \{v_1, v_2, ..., v_N\}$.
		The user-item interaction matrix ${\bf Y} \in \mathbb R^{M \times N}$ is defined according to users' implicit feedback, where $y_{uv} = 1$ indicates that user $u$ engages with item $v$, such as  clicking, browsing, or purchasing; otherwise $y_{uv} = 0$.
		Additionally, we also have a knowledge graph $\mathcal G$, which is comprised of entity-relation-entity triples $(h, r, t)$.
		Here $h \in \mathcal E$, $r \in \mathcal R$, and $t \in \mathcal E$ denote the head, relation, and tail of a knowledge triple, $\mathcal E$ and $\mathcal R$ are the set of entities and relations in the knowledge graph, respectively.
		For example, the triple (\textit{A Song of Ice and Fire}, \textit{book.book.author}, \textit{George Martin}) states the fact that George Martin writes the book ``A Song of Ice and Fire".
		In many recommendation scenarios, an item $v \in \mathcal V$ corresponds to one entity $e \in \mathcal E$.
		For example, in book recommendation, the item ``A Song of Ice and Fire" also appears in the knowledge graph as an entity with the same name.
	
		Given the user-item interaction matrix $\bf Y$ as well as the knowledge graph $\mathcal G$, we aim to predict whether user $u$ has potential interest in item $v$ with which he has had no interaction before.
		Our goal is to learn a prediction function ${\hat y}_{uv} = \mathcal F(u, v | \Theta, \bf Y, \mathcal G)$, where ${\hat y}_{uv}$ denotes the probability that user $u$ will engage with item $v$, and $\Theta$ denotes the model parameters of function $\mathcal F$.

	\subsection{KGCN Layer}
		KGCN is proposed to capture high-order structural proximity among entities in a knowledge graph.		
		We start by describing a single KGCN layer in this subsection.
		Consider a candidate pair of user $u$ and item (entity) $v$.
		We use $\mathcal N(v)$ to denote the set of entities directly connected to $v$,\footnote{The knowledge graph $\mathcal G$ is treated undirected.} and $r_{e_i, e_j}$ to denote the relation between entity $e_i$ and $e_j$.
		We also use a function $g: \mathbb R^d \times \mathbb R^d \rightarrow \mathbb R$ (e.g., inner product) to compute the score between a user and a relation:
		\begin{equation}
		\label{eq:ur_score}
			\pi_r^u = g({\bf u}, {\bf r}),
		\end{equation}
		where ${\bf u} \in \mathbb R^d$ and ${\bf r} \in \mathbb R^d$ are the representations of user $u$ and relation $r$, respectively, $d$ is the dimension of representations.
		In general, $\pi_r^u$ characterizes the importance of relation $r$ to user $u$.
		For example, a user may have more potential interests in the movies that share the same ``star" with his historically liked ones, while another user may be more concerned about the ``genre" of movies.
		
		To characterize the topological proximity structure of item $v$, we compute the linear combination of $v$'s neighborhood:
		\begin{equation}
		\label{eq:agg}
			{\bf v}_{\mathcal N(v)}^u = \sum_{e \in \mathcal N(v)} \tilde \pi_{r_{v, e}}^u {\bf e},
		\end{equation}
		where $\tilde \pi_{r_{v, e}}^u$ is the normalized user-relation score
		\begin{equation}
			\tilde \pi_{r_{v, e}}^u = \frac{\exp{(\pi_{r_{v, e}}^u})}{\sum_{e \in \mathcal N(v)} \exp{(\pi_{r_{v, e}}^u)}},
		\end{equation}
		and $\bf e$ is the representation of entity $e$.
		User-relation scores act as \textit{personalized filters} when computing an entity's neighborhood representation, since we aggregate the neighbors with bias with respect to these user-specific scores.
		
\begin{figure}
			\centering
			\begin{subfigure}[b]{0.2\textwidth}
   				\includegraphics[width=\textwidth]{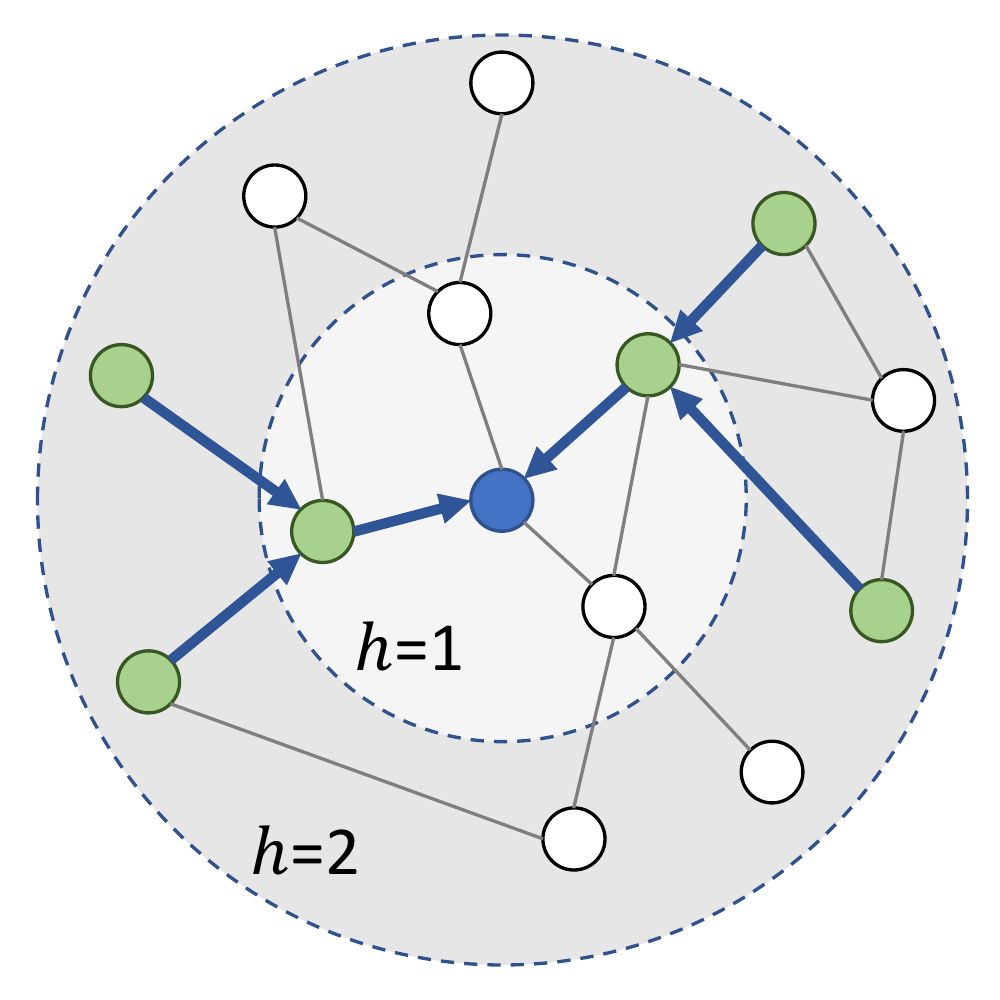}
   				\caption{}
   				\label{fig:framework1}
			\end{subfigure}
			\hfill
			\begin{subfigure}[b]{0.24\textwidth}
				\includegraphics[width=\textwidth]{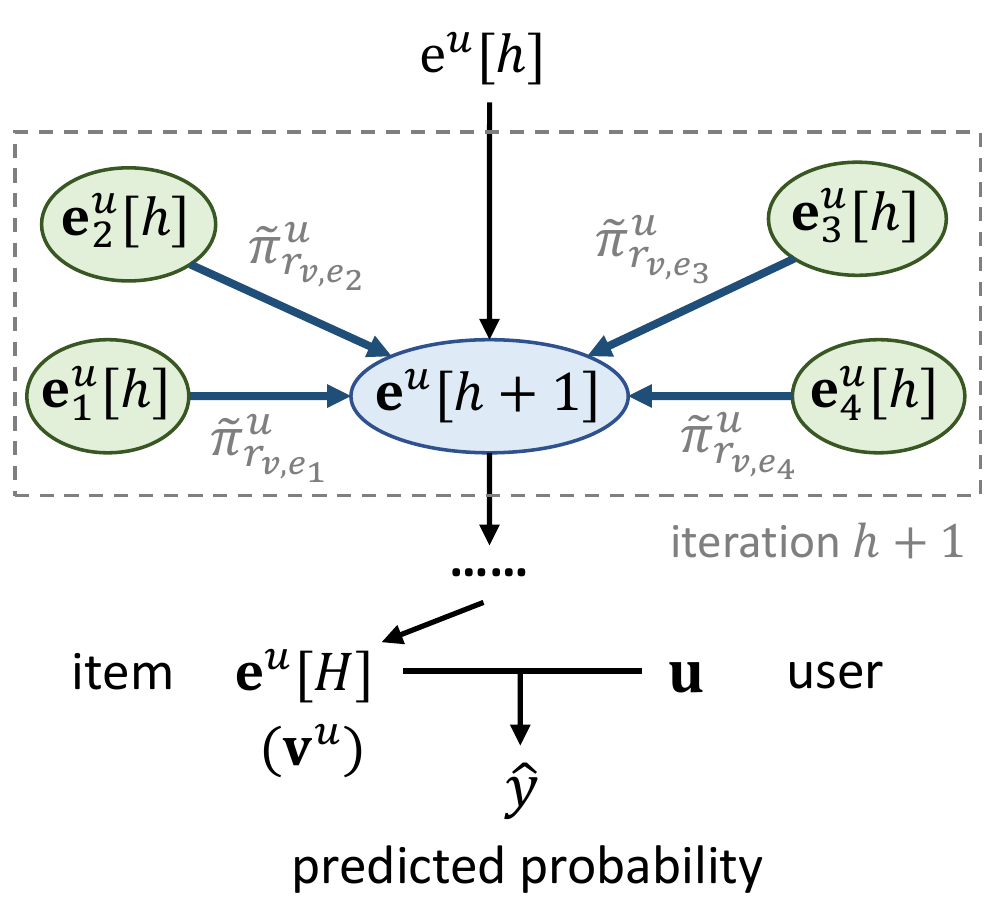}
				\caption{}
				\label{fig:framework2}
			\end{subfigure}
			\caption{(a) A two-layer receptive field (green entities) of the blue entity in a KG. (b) The framework of KGCN.}	
			\label{fig:framework}
		\end{figure}

		In a real-world knowledge graph, the size of $\mathcal N(e)$ may vary significantly over all entities.
		To keep the computational pattern of each batch fixed and more efficient, we uniformly sample a fixed-size set of neighbors for each entity instead of using its full neighbors.
		Specifically, we compute the neighborhood representation of entity $v$ as ${\bf v}_{\mathcal S(v)}^u$, where $\mathcal S(v) \triangleq \{ e \ | \ e \sim \mathcal N(v) \}$ and $| \mathcal S(v) | = K$ is a configurable constant.\footnote{Technically, $\mathcal S(v)$ may contain duplicates if $\mathcal N(v) < K$.}
		In KGCN, $\mathcal S(v)$ is also called the (single-layer) \textit{receptive field} of entity $v$, as the final representation of $v$ is sensitive to these locations.
		Figure \ref{fig:framework1} gives an illustrative example of a two-layer receptive field for a given entity, where $K$ is set as $2$.
		
		The final step in a KGCN layer is to aggregate the entity representation $\bf v$ and its neighborhood representation ${\bf v}_{\mathcal S(v)}^u$ into a single vector.
		We implement three types of aggregators $agg: \mathbb R^d \times \mathbb R^d \rightarrow \mathbb R^d$ in KGCN:
		\begin{itemize}
			\item
				\textit{Sum aggregator} takes the summation of two representation vectors, followed by a nonlinear transformation:
				\begin{equation}
					agg_{sum} = \sigma \left( {\bf W} \cdot ({\bf v} + {\bf v}_{\mathcal S(v)}^u) + {\bf b} \right),
				\end{equation}
				where $\bf W$ and $\bf b$ are transformation weight and bias, respectively, and $\sigma$ is the nonlinear function such \textit{ReLU}.
			\item
				\textit{Concat aggregator} \cite{hamilton2017inductive} concatenates the two representation vectors first before applying nonlinear transformation:
				\begin{equation}
					agg_{concat} = \sigma \left( {\bf W} \cdot concat({\bf v}, {\bf v}_{\mathcal S(v)}^u) + {\bf b} \right).
				\end{equation}
			\item
				\textit{Neighbor aggregator} \cite{velickovic2017graph} directly takes the neighborhood representation of entity $v$ as the output representation:
				\begin{equation}
					agg_{neighbor} = \sigma \left( {\bf W} \cdot {\bf v}_{\mathcal S(v)}^u + {\bf b} \right).
				\end{equation}
		\end{itemize}
		Aggregation is a key step in KGCN, because the representation of an item is bound up with its neighbors by aggregation.
		We will evaluate the three aggregators in experiments.

	\subsection{Learning Algorithm}
		Through a single KGCN layer, the final representation of an entity is dependent on itself as well as its immediate neighbors, which we name 1-order entity representation.
		It is natural to extend KGCN from one layer to multiple layers to reasonably explore users' potential interests in a broader and deeper way.
		The technique is intuitive:
		Propagating the initial representation of each entity (0-order representation) to its neighbors leads to 1-order entity representation, then we can repeat this procedure, i.e., further propagating and aggregating 1-order representations to obtain 2-order ones.
		Generally speaking, the $h$-order representation of an entity is a mixture of initial representations of itself and its neighbors up to $h$ hops away.
		This is an important property for KGCN, which we will discuss in the next subsection.
		
		\SetKwProg{Fn}{Function}{}{}
		\begin{algorithm}[t]
			\caption{KGCN algorithm}
			\label{alg:kgcn}
			\KwIn{Interaction matrix $\bf Y$; knowledge graph $\mathcal G(\mathcal E, \mathcal R)$; neighborhood sampling mapping $\mathcal S: e \rightarrow 2^{\mathcal E}$; trainable parameters: $\{{\bf u}\}_{u \in \mathcal U}$, $\{{\bf e}\}_{e \in \mathcal E}$, $\{{\bf r}\}_{r \in \mathcal R}$, $\{{\bf W}_i, {\bf b}_i\}_{i=1}^H$; hyper-parameters: $H$, $d$, $g(\cdot)$, $f(\cdot)$, $\sigma(\cdot)$, $agg(\cdot)$}
			\KwOut{Prediction function $\mathcal F(u, v | \Theta, \bf Y, \mathcal G)$}
			\While{KGCN not converge}{
				\For{$(u, v)$ in $\bf Y$}{
					$\{ \mathcal M[i] \}_{i=0}^H \leftarrow$ Get-Receptive-Field ($v$)\;
					${\bf e}^u[0] \leftarrow {\bf e}, \forall e \in \mathcal M[0]$\;
					\For{$h = 1, ..., H$}{
						\For{$e \in \mathcal M[h]$}{
							${\bf e}_{\mathcal S(e)}^u[h-1] \leftarrow \sum_{e' \in \mathcal S(e)} \tilde \pi_{r_{e, e'}}^u {\bf e'}^u[h-1]$\;
							${\bf e}^u[h] \leftarrow agg \left( {\bf e}_{\mathcal S(e)}^u[h-1], {\bf e}^u[h-1] \right)$\;
						}
					}
					${\bf v}^u \leftarrow {\bf e}^u[H]$\;
					Calculate predicted probability $\hat y_{uv} = f({\bf u}, {\bf v}^u)$\;
					Update parameters by gradient descent\;
				}
			}
			\Return $\mathcal F$\;
			\BlankLine
			\Fn{\rm{Get-Receptive-Field} ($v$)}{
				$\mathcal M[H] \leftarrow v$\;
				\For{$h = H-1, ..., 0$}{
					$\mathcal M[h] \leftarrow \mathcal M[h+1]$\;
					\For{$e \in \mathcal M[h+1]$}{
						$\mathcal M[h] \leftarrow \mathcal M[h] \cup \mathcal S(e)$\;
					}
				}
				\Return $\{ \mathcal M[i] \}_{i=0}^H$\;
			}
		\end{algorithm}
		
		The formal description of the above steps is presented in Algorithm \ref{alg:kgcn}.
		$H$ denotes the maximum depth of receptive field (or equivalently, the number of aggregation iterations), and a suffix $[h]$ attached by a representation vector denotes $h$-order.
		For a given user-item pair $(u, v)$ (line 2), we first calculate the receptive field $\mathcal M$ of $v$ in an iterative layer-by-layer manner (line 3, 13-19).
		Then the aggregation is repeated $H$ times (line 5):
		In iteration $h$, we calculate the neighborhood representation of each entity $e \in \mathcal M[h]$ (line 7), then aggregate it with its own representation ${\bf e}^u[h-1]$ to obtain the one to be used at the next iteration (line 8).
		The final $H$-order entity representation is denoted as ${\bf v}^u$ (line 9), which is fed into a function $f: \mathbb R^d \times \mathbb R^d \rightarrow \mathbb R$ together with user representation $\bf u$ for predicting the probability:
		\begin{equation}
			\hat y_{uv} = f({\bf u}, {\bf v}^u).
		\end{equation}
		
		Figure \ref{fig:framework2} illustrates the KGCN algorithm in one iteration, in which the entity representation ${\bf v}^u[h]$ and neighborhood representations (green nodes) of a given node are mixed to form its representation for the next iteration (blue node).
		
		Note that Algorithm \ref{alg:kgcn} traverses all possible user-item pairs (line 2).
		To make computation more efficient, we use a negative sampling strategy during training.
		The complete loss function is as follows:
		\begin{equation}
		\label{eq:loss}
		\small
		\begin{split}
			\mathcal L = \sum_{u \in \mathcal U} \bigg(\sum_{v: y_{uv}=1} \mathcal J(y_{uv}, \hat y_{uv}) - \sum_{i=1}^{T^u} \mathbb E_{v_i \sim P(v_i)} \mathcal J(y_{uv_i}, \hat y_{uv_i}) \bigg) + \lambda \| \mathcal F \|^2_2,
		\end{split}
		\end{equation}
		where $\mathcal J$ is cross-entropy loss, $P$ is a negative sampling distribution, and $T^u$ is the number of negative samples for user $u$.
		In this paper, $T^u = |\{ v:y_{uv}=1 \}|$ and $P$ follows a uniform distribution.
		The last term is the L2-regularizer.

\section{Experiments}
	In this section, we evaluate KGCN on three real-world scenarios: movie, book, and music recommendations.
	
	\subsection{Datasets}
		We utilize the following three datasets in our experiments for movie, book, and music recommendation, respectively:
		\begin{itemize}
			\item
				\textbf{MovieLens-20M}\footnote{https://grouplens.org/datasets/movielens/} is a widely used benchmark dataset in movie recommendations, which consists of approximately 20 million explicit ratings (ranging from 1 to 5) on the MovieLens website.
			\item
				\textbf{Book-Crossing}\footnote{http://www2.informatik.uni-freiburg.de/\textasciitilde cziegler/BX/} contains 1 million ratings (ranging from 0 to 10) of books in the Book-Crossing community.
			\item
				\textbf{Last.FM}\footnote{https://grouplens.org/datasets/hetrec-2011/} contains musician listening information from a set of 2 thousand users from Last.fm online music system.
		\end{itemize}
		
		Since the three datasets are explicit feedbacks, we transform them into implicit feedback where each entry is marked with 1 indicating that the user has rated the item positively, and sample an unwatched set marked as 0 for each user.
		The threshold of positive rating is 4 for MovieLens-20M, while no threshold is set for Book-Crossing and Last.FM due to their sparsity.
		
		We use Microsoft Satori\footnote{https://searchengineland.com/library/bing/bing-satori} to construct the knowledge graph for each dataset.
		We first select a subset of triples from the whole KG with a confidence level greater than 0.9.
		Given the sub-KG, we collect Satori IDs of all valid movies/books/musicians by matching their names with tail of triples \textit{(head, film.film.name, tail)}, \textit{(head, book.book.title, tail)}, or \textit{(head, type.object.name, tail)}.
		Items with multiple matched or no matched entities are excluded for simplicity.
		We then match the item IDs with the head of all triples and select all well-matched triples from the sub-KG.
		The basic statistics of the three datasets are presented in Table \ref{table:statistics}.
		
		\begin{table}[t]
			\centering
			\setlength{\tabcolsep}{5pt}
			\caption{Basic statistics and hyper-parameter settings for the three datasets ($K$: neighbor sampling size, $d$: dimension of embeddings, $H$: depth of receptive field, $\lambda$: L2 regularizer weight, $\eta$: learning rate).}
			\begin{tabular}{c|ccc}
				\hline
				& MovieLens-20M & Book-Crossing & Last.FM\\
				\hline
				\# users & 138,159 & 19,676 & 1,872\\
				\# items & 16,954 & 20,003 & 3,846\\
				\# interactions & 13,501,622 & 172,576 & 42,346\\
				\# entities & 102,569 & 25,787 & 9,366\\
				\# relations & 32 & 18 & 60\\
				\# KG triples & 499,474 & 60,787 & 15,518\\
				\hline
				$K$ & 4 & 8 & 8\\
				$d$ & 32 & 64 & 16\\
				$H$ & 2 & 1 & 1\\
				$\lambda$ & $10^{-7}$ & $2 \times 10^{-5}$ & $10^{-4}$\\
				$\eta$ & $2 \times 10^{-2}$ & $2 \times 10^{-4}$ & $5 \times 10^{-4}$\\
				batch size & 65,536 & 256 & 128\\
				\hline
			\end{tabular}
			\label{table:statistics}
		\end{table}
		
	\begin{table*}[t]
    	\centering
    	\setlength{\tabcolsep}{8pt}
    	\caption{The results of $AUC$ and $F1$ in CTR prediction.}
    	\begin{tabular}{c|llllll}
    		\hline
    		\multirow{2}{*}{Model} & \multicolumn{2}{c}{MovieLens-20M} & \multicolumn{2}{c}{Book-Crossing} & \multicolumn{2}{c}{Last.FM} \\
            & \multicolumn{1}{c}{\textit{AUC}} & \multicolumn{1}{c}{\textit{F1}} & \multicolumn{1}{c}{\textit{AUC}} & \multicolumn{1}{c}{\textit{F1}} & \multicolumn{1}{c}{\textit{AUC}} & \multicolumn{1}{c}{\textit{F1}} \\
            \hline
            SVD & 0.963 (-1.5\%) & 0.919 (-1.4\%) & 0.672 (-8.9\%) & 0.635 (-7.7\%) & 0.769 (-3.4\%) & 0.696 (-3.5\%) \\
            LibFM & 0.959 (-1.9\%) & 0.906 (-2.8\%) & 0.691 (-6.4\%) & 0.618 (-10.2\%) & 0.778 (-2.3\%) & 0.710 (-1.5\%) \\
            LibFM + TransE & 0.966 (-1.2\%) & 0.917 (-1.6\%) & 0.698 (-5.4\%) & 0.622 (-9.6\%) & 0.777 (-2.4\%) & 0.709 (-1.7\%) \\
            PER & 0.832 (-14.9\%) & 0.788 (-15.5\%) & 0.617 (-16.4\%) & 0.562 (-18.3\%) & 0.633 (-20.5\%) & 0.596 (-17.3\%) \\
            CKE & 0.924 (-5.5\%) & 0.871 (-6.5\%) & 0.677 (-8.3\%) & 0.611 (-11.2\%) & 0.744 (-6.5\%) & 0.673 (-6.7\%) \\
            RippleNet & 0.968 (-1.0\%) & 0.912 (-2.1\%) & 0.715 (-3.1\%) & 0.650 (-5.5\%) & 0.780 (-2.0\%) & 0.702 (-2.6\%) \\
            \hline
            KGCN-sum & \textbf{0.978} & \textbf{0.932}* & \textbf{0.738} & \textbf{0.688}* & 0.794 (-0.3\%) & 0.719 (-0.3\%) \\
            KGCN-concat & 0.977 (-0.1\%) & 0.931 (-0.1\%) & 0.734 (-0.5\%) & 0.681 (-1.0\%) & \textbf{0.796}* & \textbf{0.721}* \\
            KGCN-neighbor & 0.977 (-0.1\%) & \textbf{0.932}* & 0.728 (-1.4\%) & 0.679 (-1.3\%) & 0.781 (-1.9\%) & 0.699 (-3.1\%) \\
            \hline
            KGCN-avg & 0.975 (-0.3\%) & 0.929 (-0.3\%) & 0.722 (-2.2\%) & 0.682 (-0.9\%) & 0.774 (-2.8\%) & 0.692 (-4.0\%) \\
            \hline
		\end{tabular}
		\label{table:ctr}
		\footnotesize \flushleft{* Statistically significant improvement by unpaired two-sample $t$-test with $p=0.1$.}
	\end{table*}

	\subsection{Baselines}
		We compare the proposed KGCN with the following baselines, in which the first two baselines are KG-free while the rest are all KG-aware methods.
		Hyper-parameter settings for baselines are introduced in the next subsection.
		\begin{itemize}
			\item
				\textbf{SVD} \cite{koren2008factorization} is a classic CF-based model using inner product to model user-item interactions.\footnote{We have tried NCF \cite{he2017neural}, i.e., replacing inner product with neural networks, but the result is inferior to SVD. Since SVD and NCF are similar, we only present the better one here.}
			\item
				\textbf{LibFM} \cite{rendle2012factorization} is a feature-based factorization model in CTR scenarios.
				We concatenate user ID and item ID as input for LibFM.
			\item
				\textbf{LibFM + TransE} extends LibFM by attaching an entity representation learned by TransE \cite{bordes2013translating} to each user-item pair.
			\item
				\textbf{PER} \cite{yu2014personalized} treats the KG as heterogeneous information networks and extracts meta-path based features to represent the connectivity between users and items.
			\item
				\textbf{CKE} \cite{zhang2016collaborative} combines CF with structural, textual, and visual knowledge in a unified framework for recommendation.
				We implement CKE as CF plus a structural knowledge module in this paper.
			\item
				\textbf{RippleNet} \cite{wang2018ripple} is a memory-network-like approach that propagates users' preferences on the KG for recommendation.
		\end{itemize}

	\subsection{Experiments Setup}
		In KGCN, we set functions $g$ and $f$ as inner product, $\sigma$ as \textit{ReLU} for non-last-layer aggregator and $tanh$ for last-layer aggregator.
		Other hyper-parameter settings are provided in Table \ref{table:statistics}.
		The hyper-parameters are determined by optimizing $AUC$ on a validation set.		
		For each dataset, the ratio of training, evaluation, and test set is $6:2:2$.
		Each experiment is repeated $3$ times, and the average performance is reported.
		We evaluate our method in two experiment scenarios:
		(1) In click-through rate (CTR) prediction, we apply the trained model to predict each interaction in the test set.
		We use $AUC$ and $F1$ to evaluate CTR prediction.
		(2) In top-$K$ recommendation, we use the trained model to select $K$ items with highest predicted click probability for each user in the test set, and choose $Recall@K$ to evaluate the recommended sets.
		All trainable parameters are optimized by Adam algorithm.
		The code of KGCN-LS is implemented under Python 3.6, TensorFlow 1.12.0, and NumPy 1.14.3.
		
		The hyper-parameter settings for baselines are as follows.
		For SVD, we use the unbiased version (i.e., the predicted rating is modeled as $r_{pq} = {\bf p}^\top {\bf q}$).
		The dimension and learning rate for the four datasets are set as: $d = 8$, $\eta = 0.5$ for MovieLens-20M, Book-Crossing; $d = 8$, $\eta = 0.1$ for Last.FM.
		For LibFM, the dimension is $\{1, 1, 8\}$ and the number of training epochs is $50$.
		The dimension of TransE is $32$.
		For PER, we use manually designed user-item-attribute-item paths as features (i.e., ``user-movie-director-movie", ``user-movie-genre-movie", and ``user-movie-star-movie" for MovieLens-20M; ``user-book-author-book" and ``user-book-genre-book" for Book-Crossing, ``user-musician-date\_of\_birth-musician" (date of birth is discretized), ``user-musician-country-musician", and ``user-musician-genre-musician" for Last.FM).
		For CKE, the dimension of the three datasets are $64$, $128$, $64$.
		The training weight for KG part is $0.1$ for all datasets.
		The learning rate are the same as in SVD.
		For RippleNet, $d=8$, $H=2$, $\lambda_1 = 10^{-6}$, $\lambda_2=0.01$, $\eta=0.01$ for MovieLens-20M; $d=16$, $H=3$, $\lambda_1 = 10^{-5}$, $\lambda_2=0.02$, $\eta=0.005$ for Last.FM.
		Other hyper-parameters are the same as reported in their original papers or as default in their codes.

	\begin{figure*}[t]
		\centering
        \begin{subfigure}[b]{0.32\textwidth}
            \includegraphics[width=\textwidth]{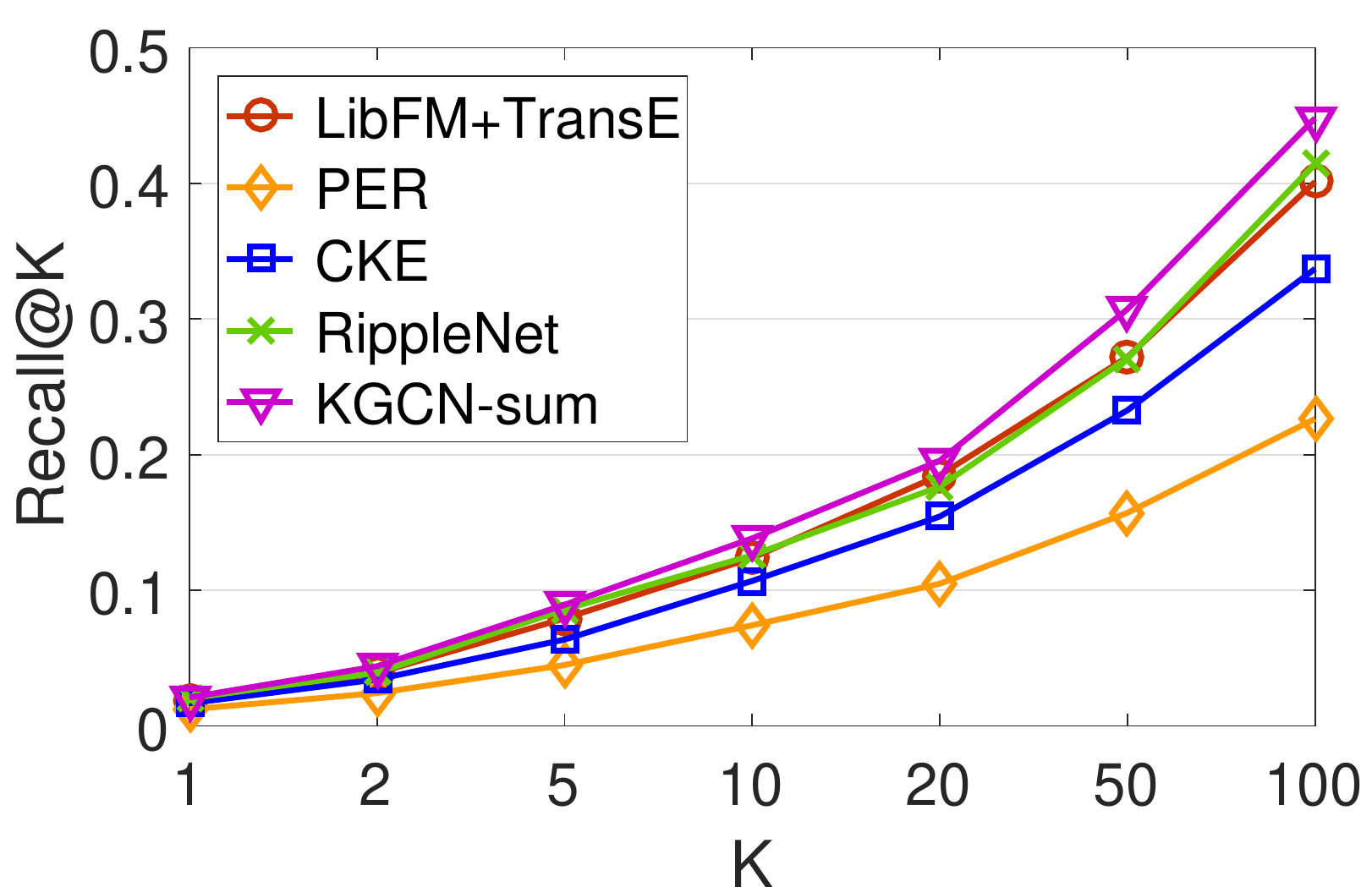}
            \caption{MovieLens-20M}
            \label{fig:res_1}
        \end{subfigure}
        \hfill
        \begin{subfigure}[b]{0.32\textwidth}
            \includegraphics[width=\textwidth]{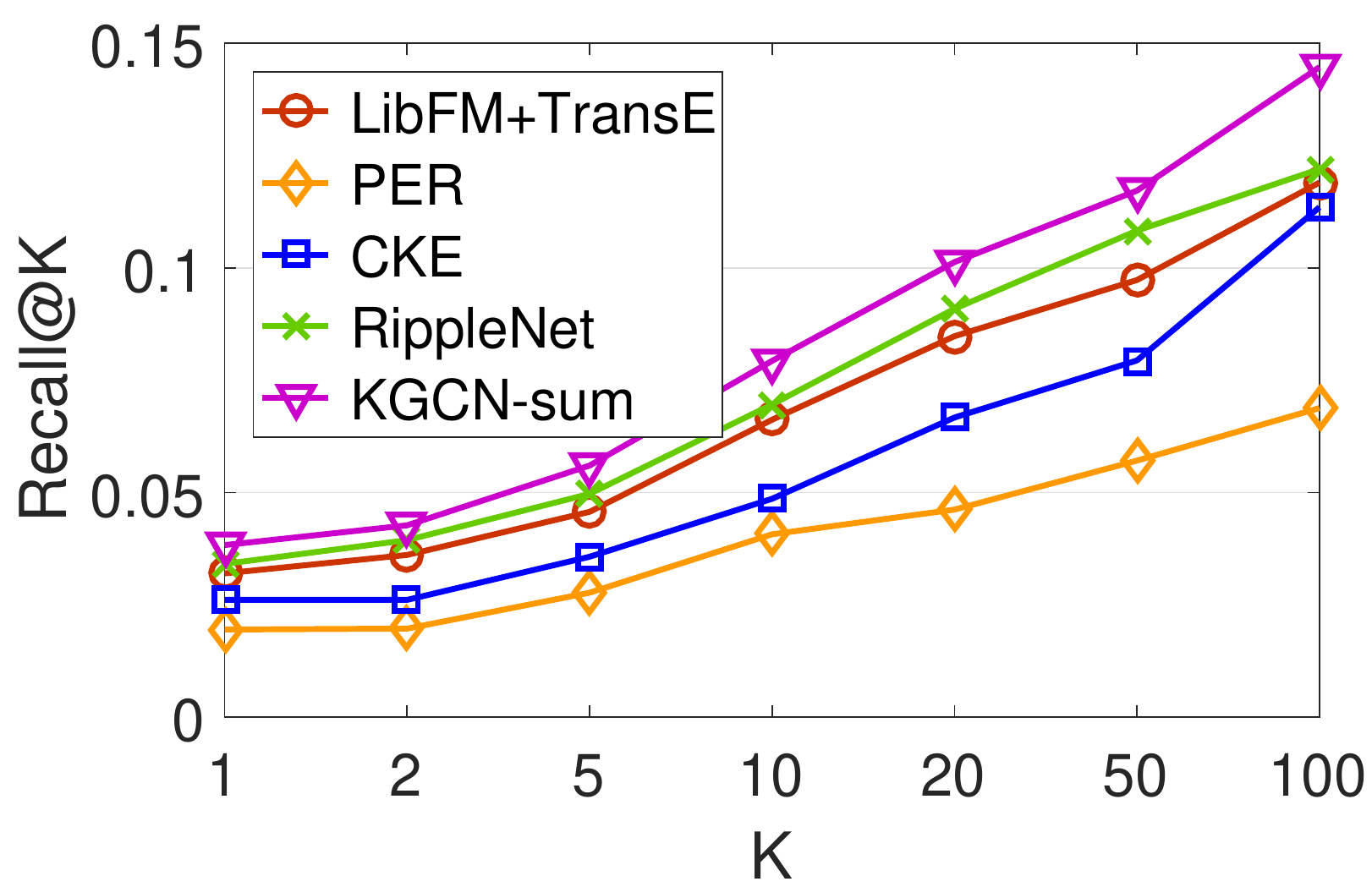}
            \caption{Book-Crossing}
            \label{fig:res_2}
        \end{subfigure}
        \hfill
        \begin{subfigure}[b]{0.32\textwidth}
            \includegraphics[width=\textwidth]{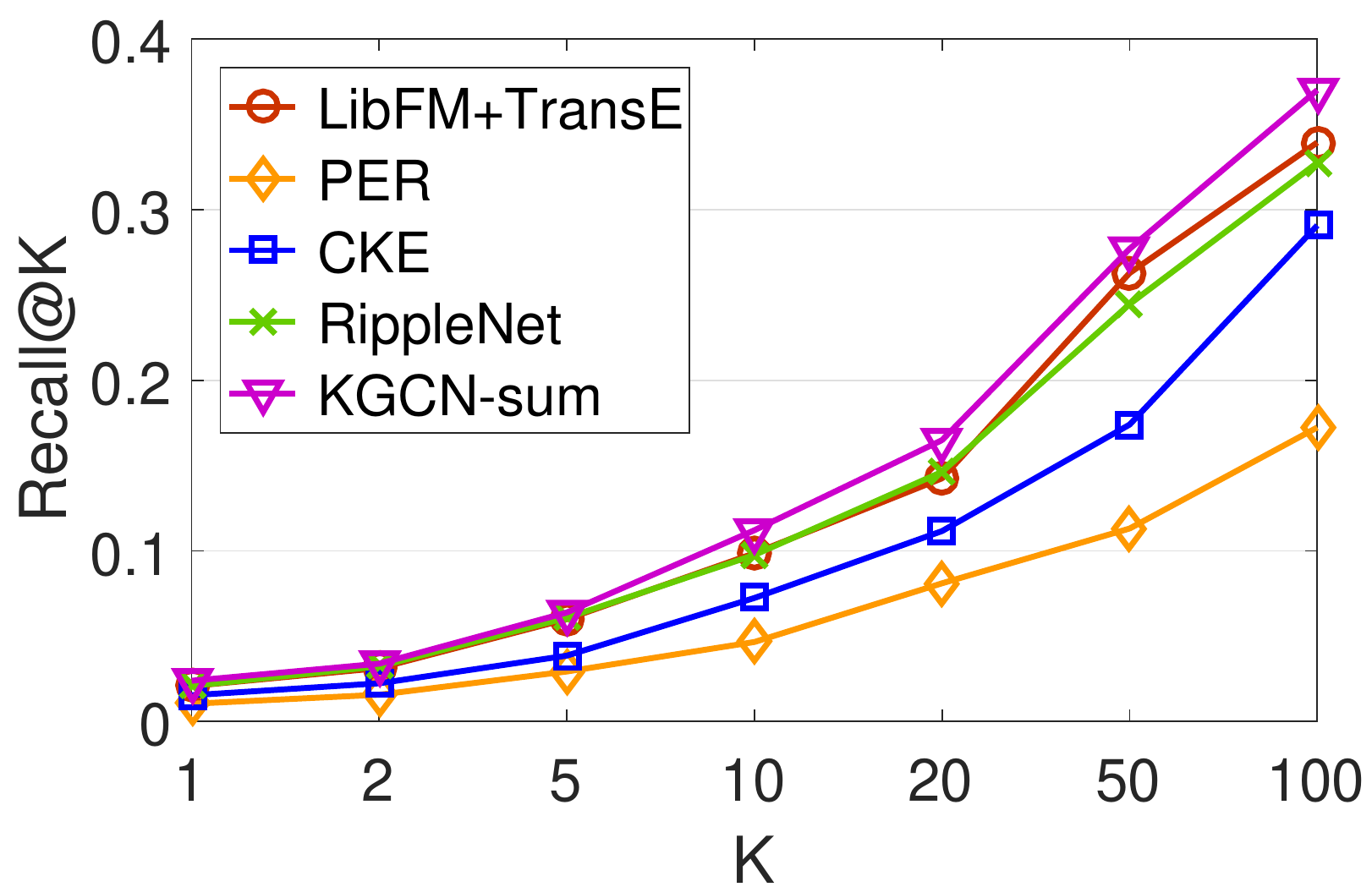}
            \caption{Last.FM}
            \label{fig:res_3}
        \end{subfigure}
        \vspace{-0.1in}
        \caption{The results of $Recall@K$ in top-$K$ recommendation.}
        \label{fig:topk}
    \end{figure*}

	\subsection{Results}
		The results of CTR prediction and top-$K$ recommendation are presented in Table \ref{table:ctr} and Figure \ref{fig:topk}, respectively (SVD, LibFM and other variants of KGCN are not plotted in Figure \ref{fig:topk} for clarity).
        We have the following observations:
        \begin{itemize}
        	\item
        		In general, we find that the improvements of KGCN on book and music are higher than movie.
        		This demonstrates that KGCN can well address sparse scenarios, since Book-Crossing and Last.FM are much sparser than MovieLens-20M.
        	\item
        		The performance of KG-free baselines, SVD and LibFM, are actually better than the two KG-aware baselines PER and CKE, which indicates that PER and CKE cannot make full use of the KG with manually designed meta-paths and TransR-like regularization.
        	\item
        		LibFM + TransE is better than LibFM in most cases, which demonstrates that the introduction of KG is helpful for recommendation in general.
        	\item
        		PER performs worst among all baselines, since it is hard to define optimal meta-paths in reality.
        	\item
        		RippleNet shows strong performance compared with other baselines.
        		Note that RippleNet also uses multi-hop neighborhood structure, which interestingly shows that capturing proximity information in the KG is essential for recommendation.
        \end{itemize}
        
        The last four rows in Table \ref{table:ctr} summarize the performance of KGCN variants.
        The first three (sum, concat, neighbor) correspond to different aggregators introduced in the preceding section, while the last variant KGCN-avg is a reduced case of KGCN-sum where neighborhood representations are directly averaged without user-relation scores (i.e., ${\bf v}_{\mathcal N(v)}^u = \sum_{e \in \mathcal N(v)} {\bf e}$ instead of Eq. (\ref{eq:agg})).
        Therefore, KGCN-avg is used to examine the efficacy of the ``attention mechanism".
        From the results we find that:
        \begin{itemize}
        	\item
        		KGCN outperforms all baselines by a significant margin, while their performances are slightly distinct:
        		KGCN-sum performs best in general, while the performance of KGCN-neighbor shows a clear gap on Book-Crossing and Last.FM.
        		This may be because the neighbor aggregator uses the neighborhood representation only, thus losing useful information from the entity itself.
        	\item
        		KGCN-avg performs worse than KGCN-sum, especially in Book-Crossing and Last.FM where interactions are sparse.
        		This demonstrates that capturing users' personalized preferences and semantic information of the KG do benefit the recommendation.
        \end{itemize}

		\subsubsection{Impact of neighbor sampling size.}
			\begin{table}[t]
				\centering
				\setlength{\tabcolsep}{4pt}
				\caption{$AUC$ result of KGCN with different neighbor sampling size $K$.}
				\begin{tabular}{c|cccccc}
					\hline
					$K$ & 2 & 4 & 8 & 16 & 32 & 64\\
					\hline
					MovieLens-20M & 0.978 & \textbf{0.979} & 0.978 & 0.978 & 0.977 & 0.978\\
					Book-Crossing & 0.680 & 0.727 & \textbf{0.736} & 0.725 & 0.711 & 0.723\\
					Last.FM & 0.791 & 0.794 & \textbf{0.795} & 0.793 & 0.794 & 0.792\\
					\hline
				\end{tabular}
				\label{table:K}
			\end{table}	
			
			We vary the size of sampled neighbor $K$ to investigate the efficacy of usage of the KG.
			From Table \ref{table:K} we observe that KGCN achieves the best performance when $K=4$ or $8$.
			This is because a too small $K$ does not have enough capacity to incorporate neighborhood information, while a too large $K$ is prone to be misled by noises.
			
		\subsubsection{Impact of depth of receptive field.}
			\begin{table}[t]
				\centering
				\setlength{\tabcolsep}{8pt}
				\caption{$AUC$ result of KGCN with different depth of receptive field $H$.}
				\begin{tabular}{c|cccc}
					\hline
					$H$ & 1 & 2 & 3 & 4\\
					\hline
					MovieLens-20M & 0.972 & \textbf{0.976} & 0.974 & 0.514\\
					Book-Crossing & \textbf{0.738} & 0.731 & 0.684 & 0.547\\
					Last.FM & \textbf{0.794} & 0.723 & 0.545 & 0.534\\
					\hline
				\end{tabular}
				\label{table:H}
			\end{table}
			
			We investigate the influence of depth of receptive field in KGCN by varying $H$ from 1 to 4.
			The results are shown in Table \ref{table:H}, which demonstrate that KGCN is more sensitive to $H$ compared to $K$.
			We observe the occurrence of serious model collapse when $H=3$ or $4$, as a larger $H$ brings massive noises to the model.
			This is also in accordance with our intuition, since a too long relation-chain makes little sense when inferring inter-item similarities.
			An $H$ of 1 or 2 is enough for real cases according to the experiment results.

		\subsubsection{Impact of dimension of embedding.}
			\begin{table}[t]
				\centering
				\setlength{\tabcolsep}{4pt}
				\caption{$AUC$ result of KGCN with different dimension of embedding.}
				\begin{tabular}{c|cccccc}
					\hline
					$d$ & 4 & 8 & 16 & 32 & 64 & 128\\
					\hline
					MovieLens-20M & 0.968 & 0.970 & 0.975 & \textbf{0.977} & 0.973 & 0.972\\
					Book-Crossing & 0.709 & 0.732 & 0.733 & 0.735 & \textbf{0.739} & 0.736\\
					Last.FM & 0.789 & 0.793 & \textbf{0.797} & 0.793 & 0.790 & 0.789\\
					\hline
				\end{tabular}
				\label{table:d}
			\end{table}
			
			Lastly, we examine the influence of dimension of embedding $d$ on performance of KGCN.
			The result in Table \ref{table:d} is rather intuitive:
			Increasing $d$ initially can boost the performance since a larger $d$ can encode more information of users and entities, while a too large $d$ adversely suffers from overfitting.

\section{Conclusions and Future Work}
	This paper proposes knowledge graph convolutional networks for recommender systems.
	KGCN extends non-spectral GCN approaches to the knowledge graph by aggregating neighborhood information selectively and biasedly, which is able to learn both structure information and semantic information of the KG as well as users' personalized and potential interests.
	We also implement the proposed method in a minibatch fashion, which is able to operate on large datasets and knowledge graphs.
	Through extensive experiments on real-world datasets, KGCN is shown to consistently outperform state-of-the-art baselines in movie, book, and music recommendation.
	
	We point out three avenues for future work.
	(1) In this work we uniformly sample from the neighbors of an entity to construct its receptive field.
	Exploring a non-uniform sampler (e.g., importance sampling) is an important direction of future work.
	(2) This paper (and all literature) focuses on modeling item-end KGs.
	An interesting direction of future work is to investigate whether leveraging user-end KGs is useful in improving the performance of  recommendation.
	(3) Designing an algorithm to well combine KGs at the two ends is also a promising direction.

\bibliographystyle{ACM-Reference-Format}
\bibliography{reference} 

\end{document}